\documentclass[twocolumn,showpacs,floatfix,aps,nofootinbib]{revtex4}
\usepackage{epsfig}

\usepackage{graphicx}

\begin{document}

\title{The standard model on a domain-wall brane?}
\author{Rhys Davies}\email{r.davies1@physics.ox.ac.uk}
\author{Damien P. George}\email{d.george@physics.unimelb.edu.au}
\affiliation{School of Physics, Research Centre for High Energy
Physics, The University of Melbourne, Victoria 3010, Australia}
\author{Raymond R. Volkas}\email{raymondv@unimelb.edu.au}
\affiliation{School of Physics, Research Centre for High Energy
Physics, The University of Melbourne, Victoria 3010, Australia}

\begin{abstract}
We propose a $4+1$-dimensional action that is a candidate for realising a standard-model-like effective theory 
for fields dynamically localised to a domain-wall brane.  Our construction
is in part based on the conjecture that the Dvali-Shifman mechanism for dynamically
localising gauge bosons to a domain wall works correctly in $4+1$-d.  Assuming this to be so, we
require the gauge symmetry to be SU(5) in the bulk, spontaneously breaking to 
SU(3)$\otimes$SU(2)$\otimes$U(1) inside the domain wall, thus dynamically localising the standard model
gauge bosons provided that the SU(5) theory in the bulk exhibits confinement.
The wall is created jointly by a real singlet-Higgs field $\eta$ 
configured as a kink, and an SU(5) adjoint-Higgs field $\chi$ that takes nonzero values 
inside the wall. Chiral $3+1$-dimensional quarks and leptons are confined and split 
along the bulk direction via their Yukawa couplings to $\eta$ and $\chi$.  
The Higgs doublet and its colour triplet SU(5) partner are similarly localised and split.  
The splittings can suppress coloured-Higgs-induced proton decay and, because of the different 
localisation profiles, the usual SU(5) mass relation $m_e = m_d$ does not arise.
Localised gravity is generated via the Randall-Sundrum alternative to compactification.

\end{abstract}


\maketitle

\section{Introduction}

There is no known fundamental principle requiring spacetime to be $3+1$-dimensional, so
extra dimensions of space might exist.  If so, then the effective $3+1$-dimensionality we observe in
everyday life and in high-energy experiments has to be explained.  It could be that
the extra dimensions are topologically compact and small, as per the Kaluza-Klein idea.
Alternatively, the extra dimensions could be large but as yet unobserved because standard model (SM)
fields are confined to a $3+1$-d brane.  Large extra dimensions might be compact,
as proposed by Arkani-Hamed, Dimopoulos and Dvali~\cite{ADD}, or infinite, as shown in the
second of the Randall-Sundrum papers of 1999~\cite{RS2} (hereinafter RS2).  See
also Refs.~\cite{RS1, Antoniadis, Antoniadisetal, Akama, Visser}.

The purpose of this paper is to propose a $4+1$-dimensional action that is a candidate for 
realising a standard-model-like effective theory for fields dynamically localised to a domain-wall (DW) brane.
Like RS2, there is one extra dimension and it is infinite.  Unlike RS2, the brane is not a
fundamental object but rather a solitonic solution of the theory, as per the Rubakov and 
Shaposhnikov~\cite{Rub&Shap} proposal that we might live on a domain wall.
Our construction assembles a number
of dynamical localisation mechanisms into what we hope is a complete theory of a
domain-wall-localised SM.  These mechanisms are:
\begin{itemize}
\item the localisation of $3+1$-d chiral fermion zero modes through the Yukawa coupling
of $4+1$-d fermions to the background scalar fields;
\item the localisation of a SM Higgs doublet to the DW through its Higgs potential
couplings to the DW-forming scalar fields;
\item the localisation of SM gauge bosons via the Dvali-Shifman mechanism, instituted 
through a bulk that respects SU(5) gauge invariance~\cite{Georgi&Glashow};
\item the DW generalisation of the RS2 mechanism for localising gravitons.
\end{itemize}
Three of these four mechanisms involve well-established phenomena.  The Dvali-Shifman (DS)
gauge boson localisation idea remains an interesting conjecture in the $4+1$-d context, not
as yet proven to work.  What we shall show in this paper is that {\it if} one takes the
DS mechanism to work in $4+1$-d, {\it then} the construction of a DW-localised standard model
follows readily, and even elegantly.  We hope that our model spurs rigorous
studies of the DS mechanism in $4+1$-d, to either confirm it or disprove it.  Were it to be confirmed,
then our model-building setup would provide a clear pathway to the construction of phenomenologically-realistic effective
theories of DW-localised fields. We shall review the DS mechanism below.

The main aesthetic motivation for our model is to treat all spatial dimensions on an equal footing
in the action.  In particular, all these dimensions are infinite, as in the RS2 setup.  But
``dimensional democracy'' is taken further than in RS2, because that theory has
translational invariance along the extra dimension explicitly broken through the introduction of an
infinitely-thin fundamental brane into the action.  To achieve dimensional democracy we must have
no explicit brane terms in the action, but
replace the RS2 fundamental brane with a finite-thickness stable domain-wall
configuration of scalar fields.

We shall argue that our theory is likely to be the minimal way to get a purely
field-theoretic realisation of a DW-confined SM.  It is interesting that in order to achieve
this the Dvali-Shifman mechanism immediately motivates
an extension to SU(5).  We are also encouraged by the fact that some of the usual problems of
SU(5) grand unification have solutions automatically provided by the minimal theory,
without ``epicyclic'' {\it ad-hoc} fixes.  As we shall
explain below, the usual $m_d = m_e$ style SU(5) mass relations are simply absent, because the
fermion localisation realises a modified version of the split-fermion idea of Arkani-Hamed and 
Schmaltz~\cite{A-H&Schmaltz} (see also~\cite{Coulthurst}).   The down-type quarks 
necessarily have different bulk profile functions
from the charged leptons, and because the $3+1$-d masses are computed from overlap integrals of
profile functions, the quark-lepton mass degeneracy just does not arise.  The fermion splitting can
also suppress coloured-Higgs-induced proton decay.  An important loose end is that we are not
yet able to analyse gauge coupling constant unification in our unusual version of SU(5).  We shall 
explain below why a full unification study is premature.

Our focus in this paper is on model-building rather than detailed phenomenology.  We wish to
explain the logic of our construction, and provide evidence that it has good phenomenology without
supplying absolute proof.

We review the Dvali-Shifman mechanism in the next section, describe our model in the
following section, and conclude in the last section.

\section{Dvali-Shifman mechanism}
\label{sec:DS}

The most plausible mechanism for localising gauge bosons to a DW in such a way
as to preserve gauge invariance is that proposed by Dvali and Shifman~\cite{DvaliShifman}.
This requires a confining non-Abelian gauge theory in the bulk, with the symmetry G broken to a subgroup
group H inside the DW.  Massless gauge bosons corresponding to H
are then localised to the wall.  As we wish to localise the standard model fields,
the minimal choice is to take G=SU(5) and H=SU(3)$\otimes$SU(2)$\otimes$U(1).

The truth of the DS mechanism rests on quite a firm foundation for DWs residing in a background
$3+1$-d spacetime~\cite{Rubrev, RubaDub, A-H&Schmaltz2}.  Following DS, let us consider the simple toy example of G=SU(2) and H=U(1).  Place
a U(1) source charge inside the wall.  Because the SU(2)-respecting bulk is in confinement phase,
the electric field lines of the source charge cannot penetrate into the bulk.  Instead, the field lines
are repelled from the DW-bulk interface thus reducing the effective dimensionality of the Coulomb field
by one.  Adopting the 't Hooft-Mandelstam proposal that confinement arises from the magnetic dual of
superconductivity, the repulsion of field lines from the interface is readily understood from the
dual Meissner effect~\cite{RubaDub, A-H&Schmaltz2}.  

Now place the source charge in the bulk.  By confinement, which is tantamount to the expulsion of electric fields,
the electric flux from the source must form a flux tube that ends on the domain wall~\cite{RubaDub, A-H&Schmaltz2}.  
Once inside the wall the
field lines are able to spread out in the plane of the wall.  It is as if the charge was actually inside the wall:
the electric field configuration is the same at large distances inside the wall irrespective of the position
of the source.  In the quantal situation where the position of a source charge is indefinite, it follows that the
long range Coulomb field is independent of how the wave-function depends on the coordinate perpendicular to the wall
(the ``extra'' dimension).  We shall be using this result below when we assume that gauge-universality for H
holds {\it independently of the bulk profiles of the trapped fields}.

If H is non-Abelian, then these arguments generalise to the case of chromoelectric field line expulsion from the bulk.

Another perspective on the localisation physics is provided by the mass gap~\cite{DvaliShifman}.  
In the bulk, becaue of confinement,
the gauge bosons of H cannot themselves propagate but instead form constituents of propagating G glueballs.  
But the glueballs of G are massive.  In the G=SU(2) and
H=U(1) example, the U(1) gauge boson which is both massless and free inside the wall, must somehow incorporate
itself into a massive SU(2) glueball if it propagates into the bulk.  But the mass gap implies an energy cost
in doing so, thus any U(1) gauge boson inside the wall is dynamically constrained to remain there.  If H has
non-Abelian factors that are themselves in confinement phase inside the wall, then the mass gap suppression corresponds
to the H glueballs inside the wall being less massive than the G glueballs in the bulk.

These arguments are rather convincing because they rest on the well-established confinement property for
asymptotically-free non-Abelian gauge theories in $3+1$-d.  In the $4+1$-d case, the DS mechanism is a conjecture, because
$4+1$-d confinement (or lack thereof) is not properly understood.  
The main issue is that pure Yang-Mills theory is not renormalisable in $4+1$-d (or larger).
At the level of lattice gauge calculations, this corresponds to the lack of a physical
limit when taking the lattice spacing to zero.  To expand on this point, it is known that
$4+1$-d SU(2) has a first order phase transition for finite lattice spacing~\cite{Creutz}.
We have verified this conclusion for $4+1$-d SU(5) and so presumably SU(5) has a confining
phase for sufficiently large values of the gauge coupling constant.  This analysis cannot
be extended to the continuum limit, and so we must be content with $4+1$-d SU(5) exhibiting
confinement {\it below} a relevant cutoff of the theory.  Thus we consider $4+1$-d DS to be
an effective mechanism, valid below this cutoff, which does the job of confining gauge
fields to the DW.  As we remark below, any field theoretic brane-world model is non-renormalisable
and hence must be defined with an ultraviolet cutoff, so in our context we do not need to
take the continuum limit.

To the best of our knowledge, the DS mechanism has not been directly checked in $4+1$-d, which would
require more than just an analysis of the phase structure of pure Yang-Mills theory.  But we are
encouraged by lattice gauge calculations in $2+1$-d~\cite{Laine} which do verify the
mechanism.  We shall assume that it works also in $4+1$-d, and show that realistic model
building is then quite possible.

\section{The model}

We now describe our model.  As stated above, the DS mechanism immediately motivates that the bulk should respect
at least an SU(5) gauge symmetry.  By one definition of ``minimal'', the bulk symmetry should be exactly SU(5) and it should also be
the symmetry of the action; the model presented below has these features.  (It is also interesting to consider models not
adhering to these strictures.  For example, Ref.~\cite{DavidsonE6} 
describes a theory where the symmetry of the action is larger than the symmetry of the bulk.)

The SU(5) $4+1$-d field content is:
\begin{eqnarray}
{\rm scalars}:& \quad & \eta \sim 1,\ \chi \sim 24,\ \Phi \sim 5^* \nonumber\\
{\rm fermions}: & \quad & \Psi_5 \sim 5^*,\ \Psi_{10} \sim 10,
\label{eq:fields}
\end{eqnarray}
plus gauge fields.  The field $\eta$ is real, $\chi$ is conveniently represented
as a $5 \times 5$ Hermitian traceless matrix, while $\Phi$ is a fivefold column vector
of complex fields.  Chirality does not exist in $4+1$-d, so both the
$\Psi$'s are Dirac fields, with $\Psi_{10}$ being a $5\times 5$ antisymmetric matrix.
The SU(5) transformations are: $\chi \to U \chi U^{\dagger}$, $\Phi \to U^* \Phi$,
$\Psi_5 \to U^* \Psi_5$ and $\Psi_{10} \to U \Psi_{10} U^T$.
We shall, for simplicity, consider only one quark-lepton family here,
though the generalisation to three families is straightforward.  The neutrino mass
question is also deferred to later work.

Let us begin by ignoring gravity, to focus on the purely particle-physics aspects of the model.
Later we discuss what remains the same, and what changes, when RS2-style 
warped gravity is added.  
The action is
\begin{equation}
S = \int d^5x \left( T - Y_{DW} - Y_5 - V \right),
\label{eq:S}
\end{equation}
where $T$ contains the SU(5) gauge-covariant kinetic-energy terms, $Y_{DW}$ has the
Yukawa couplings of the fermions to $\eta$ and $\chi$,
\begin{eqnarray}
Y_{DW} & = & h_{5\eta} \overline{\Psi}_5 \Psi_5 \eta +  h_{5\chi} \overline{\Psi}_5 \chi^T \Psi_5 \nonumber\\
& + & h_{10\eta} {\rm Tr}(\overline{\Psi}_{10} \Psi_{10})\eta 
- 2 h_{10\chi} {\rm Tr}(\overline{\Psi}_{10} \chi \Psi_{10}),
\label{eq:YDW}
\end{eqnarray}
and $Y_5$ is the SU(5) Yukawa Lagrangian used to generate quark and lepton
masses:
\begin{equation}
Y_5 = h_{-} \overline{(\Psi_5)^c} \Psi_{10} \Phi + h_{+} (\epsilon \overline{(\Psi_{10})^c} \Psi_{10} \Phi^*) + h.c.
\end{equation}
The last term can only be written in SU(5) index notation: $\epsilon^{ijklm} {\overline{(\Psi_{10})^c}}_{ij} {\Psi_{10}}_{kl} \Phi^{*}_m$.

The Higgs potential is $V = V_{\eta\chi} + V_{\rm rest}$ where
\begin{eqnarray}
V_{\eta\chi} & = & (c\eta^2 - \mu_\chi^2){\rm Tr}(\chi^2) + a \eta {\rm Tr}(\chi^3) + \lambda_1 \left[{\rm Tr}(\chi^2)\right]^2\nonumber\\
& + & \lambda_2 {\rm Tr}(\chi^4) + l(\eta^2 - v^2)^2; 
\label{eq:Vetachi} \\
V_{\rm rest} & = & \mu_\Phi^2 \Phi^{\dagger} \Phi + \lambda_3 (\Phi^{\dagger} \Phi)^2
+ \lambda_4 \Phi^{\dagger} \Phi \eta^2
+ 2 \lambda_5 \Phi^{\dagger} \Phi {\rm Tr}(\chi^2)\nonumber\\
& + & \lambda_6 \Phi^{\dagger} (\chi^T)^2 \Phi + \lambda_7 \Phi^{\dagger} \chi^T \Phi \eta.
\label{eq:Vrest}
\end{eqnarray}
The action is invariant under the reflection discrete symmetry $y \to -y$, $\eta \to -\eta$, $\chi \to -\chi$ and
$\Psi_{5,10} \to i\Gamma^5 \Psi_{5,10}$. The $4+1$-d Dirac matrices are $\Gamma^M = (\gamma^{\mu},-i\gamma^5)$,
where $M,N = (0,1,2,3,5)$, $\mu,\nu = (0,1,2,3)$ and $x^5 \equiv y$.

The theory is non-renormalisable in $4+1$-d.  As usual in these kinds of models, there is an implicitly assumed
ultraviolet cut-off $\Lambda_{\rm UV}$ and an ultraviolet completion above that scale.  We shall adopt the agnostic
stance for both the existence and nature of this UV completion.  Our action is perhaps best considered as 
the set of lowest-dimensional operators, consistent with the stated symmetries, of a non-renormalisable
effective theory that is putatively to be derived from the UV completion.

The background DW is found by solving the $(\eta,\chi)$ Euler-Lagrange equations for 
an $x^\mu$-independent configuration obeying the boundary conditions
\begin{equation}
\eta(y=\pm\infty) = \pm v,\ \chi(y=\pm\infty) = 0,
\label{eq:bcs}
\end{equation}
corresponding to degenerate global minima of $V_{\eta\chi}$.  The spontaneously broken reflection
symmetry ensures topological stability for the DW.  Numerical solutions exist for a significant region of
parameter space.  Purely for the sake of giving a concrete example, we can impose the parameter conditions
\begin{equation}
2 \mu_\chi^2 ( c - \tilde{\lambda} ) + ( 2 c \tilde{\lambda} - 4 l \tilde{\lambda}
- c^2 ) v^2 = 0,\quad a = 0,
\end{equation}
with $\tilde{\lambda} \equiv \lambda_1 + 7 \lambda_2/30$, permitting the analytic solution
\begin{equation}
\eta(y) = v\tanh(ky),\ \chi_1(y) = A\, {\rm sech}(ky),
\label{eq:DW}
\end{equation}
where $k^2 = c v^2 - \mu_\chi^2$, $A^2 = (2 \mu_\chi^2 - c v^2)/\tilde{\lambda}$, and $\chi_1$ is the adjoint
component associated with the weak-hypercharge generator
${\rm diag}(2/3,2/3,2/3,-1,-1)\sqrt{3}/2\sqrt{5}$.  All other $\chi$ components vanish.  The configuration
$\eta$ is the usual kink, while $\chi_1$ induces SU(5) $\to$ SU(3)$\otimes$SU(2)$\otimes$U(1)
within the DW which has width $1/k$.  This background solution creates the brane, and simultaneously
confines SM gauge fields to it provided the Dvali-Shifman mechanism works with this kind of an SU(5) bulk in $4+1$-d. 

We have checked numerically that configurations such as
Eq.~(\ref{eq:DW}) are perturbatively stable against the formation of additional nonzero $\chi$ components.
The other fields, $\Phi$, $\Psi_5$ and $\Psi_{10}$, propagate in this background.  Within the wall,
the SU(5) confinement dynamics are suppressed so we can analyse classical localisation solutions for fermions and scalars 
in the usual way.  Outside the wall, the nonperturbative SU(5) physics makes calculating
impossible absent a dedicated lattice program.  Since the localisation takes place with a
characteristic distance scale of $1/k$, ignoring the nonperturbative corrections is approximately valid.

It may be worthwhile to expand on this point.
The computation of localised lowest-energy modes, such as fermion zero modes, is but the start of a systematic mode
analysis whereby $4+1$-d fields are reinterpreted as infinite towers of $3+1$-d fields (generalisation of a
Kaluza-Klein decomposition); see, for example, Refs.\cite{George, DavGeo} for an introduction to this procedure.
Schematically, one employs a mode decomposition of the form $\Psi(x,y) = \sum_n f_n(y) \psi_n(x)$ ($x \equiv x^\mu$) where the sum
is over suitable modes and includes an integration if the modes contain a continuum.  The $\psi_n$ are the
$3+1$-d fields, and the $f_n$ are mode functions.  The mode functions are usually chosen to obey certain suitable differential
equations so that the $\psi_n$ fields are those of definite mass in the effective $3+1$-d theory (in the
familiar Kaluza-Klein case of a circular extra dimension, the mode functions are chosen to be sinusoidal for
precisely this reason).  However, from a mathematical point of view the set of mode functions is just some complete
set of functions that permits the decomposition of $\Psi(x,y)$ without loss of generality, and so one has the
usual freedom to change basis by changing the mode-function set.  This is a pertinent observation for theories
that employ the non-perturbative quantum-field-theoretic Dvali-Shifman mechanism.  In the bulk, the $\psi_n$
component fields are subject to these dynamics unless $\Psi(x,y)$ is a gauge singlet, and thus the physical meaning
ascribed to the mode functions has to take this into account.  There is no problem in using the same mode-decomposition
one would use in the absence of the non-perturbative bulk, because that is simply a mathematically-valid recasting of
$\Psi(x,y)$ as an infinite set of $\psi_n$ components.  If the bulk is indeed in confinement phase,
then the gauge non-singlet $\psi_n$ fields will not propagate as free particles, so their physical interpretation will
be as constituent particles.  This is conceptually no different from expressing the QCD Lagrangian in terms of
quarks and gluons even though the propagating states are hadrons.  Fortunately we are mainly interested in the lowest
modes, whose mode functions are sharply peaked inside the domain wall, and so to a first approximation we need not
be concerned with interpretive complications because of the non-perturbative bulk.

The $4+1$-d fermions couple to the background $y$-dependent scalar fields as per $Y_{DW}$.  A full mode
decomposition analysis would involve writing each $4+1$-d fermion field as
\begin{equation}
\Psi(x,y) = \sum_m \left[ f_L^m(y) \psi_L^m(x) + f_R^m(y) \psi_R^m(x) \right],
\end{equation}
substituting this into $4+1$-d Dirac equation 
\begin{equation}
i\Gamma^M \partial_M \Psi - b(y) \Psi = 0,
\end{equation}
where $b(y)$ is given by the relevant background domain-wall scalar field configuration (see below),
and requiring that the $\psi^m$ components satisfy the
$3+1$-d Dirac equations,
\begin{equation}
i \gamma^\mu \partial_\mu \psi_{L,R}^m = m \psi_{R,L}^m.
\end{equation}
The mode functions $f^m_{L,R}$ then obey the Schr\"{o}dinger-like equations
\begin{equation}
-{f^m_{L,R}}'' + W_{\mp} f^m_{L,R} = m^2 f^m_{L,R}
\end{equation}
with effective potentials
\begin{equation}
W_{\mp}(y) = b(y)^2 \mp b'(y).
\label{eq:fermioneffectivepotential}
\end{equation}
The nature of the mode functions is then readily deduced from the analogy with the equivalent quantum-mechanical problem.
In particular, note that as $|y| \to \infty$ the potentials tend to the positive constant $b(\pm\infty)^2$.  We shall return
to this observation later on when we consider gravity.

\begin{figure}
\centering
\includegraphics[width=0.43\textwidth]{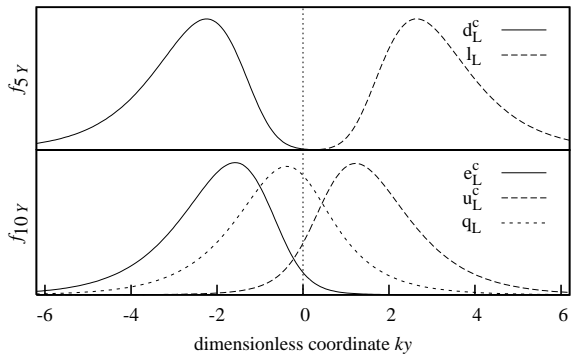}
\caption{
\small
Typical extra-dimensional profiles $f_{nY}(y)$ for the fermions
contained in the $5^*$ (top) and the $10$ (bottom).  The fields
$\eta$ and $\chi$ are as per Eq.~(\ref{eq:DW}) and parameter
choices are: $v=A=1$, $h_{n\eta}=1$, $h_{5\chi}=6$,
$h_{10\chi}=1$.  The profiles are normalised such that
$\int dy f^2_{nY}(y)=1$.
}
\label{fig:splitferm}
\end{figure}

To analyse the localisation of the lowest mode (the $m=0$ zero mode) for each fermion, the full mode analysis above is unnecessary.  
Instead, it suffices to 
solve the Dirac equations with separated variable configurations
$\Psi(x,y) = f(y) \psi_L(x)$ where the $\psi_L(x)$ are $3+1$-d
zero-mode left-chiral fields.\footnote{As pointed out by Dvali and Shifman~\cite{DvaliShifman}, as well as localising
gauge bosons the confining bulk can localise gauge non-singlet fermions and scalar fields.  However, for our application,
we have to retain the seemingly redundant localisation-to-a-kink mechanism.  The DS mechanism on its 
own will not suffice, because it will localise vector-like fermions not massless chiral fermions.  The kink
configuration is necessary for the spontaneous generation of chirality in the $3+1$-d effective theory.}
The existence of the $\chi$ Yukawa terms means that {\it different}
background fields are felt by the various SM components of $\Psi_5$ and $\Psi_{10}$.
The Dirac equations are
\begin{equation}
 \left[ i\Gamma^M \partial_M - h_{n\eta} \eta(y) - \sqrt{\frac{3}{5}}\frac{Y}{2}\, h_{n\chi} \chi_1(y) 
\right] \Psi_{nY}(x,y) = 0
\label{eq:Dirac}
\end{equation}
where $n=5,10$ and $Y$ is the weak-hypercharge of the SM components denoted $\Psi_{5Y}$ and $\Psi_{10Y}$.
{\it The SU(5) structure automatically gives different localisation points and profiles to the
different SM components -- splitting \cite{A-H&Schmaltz, Coulthurst} -- depending on 
hypercharge and whether they are in the $5^*$ or the $10$.}
The zeroes of 
\begin{equation}
b_{nY}(y) \equiv h_{n\eta} \eta(y) + \sqrt{\frac{3}{5}}\frac{Y}{2}\, h_{n\chi} \chi_1(y)
\label{eq:bnY}
\end{equation}
are the localisation centres, with the bulk profiles
\begin{equation}
f_{nY}(y) \propto e^{-\int^y b_{nY}(y') dy'}.
\label{eq:fnY}
\end{equation}
To localise $3+1$-d left-chiral fields, all the $b_{nY}$ must pass through zero with positive slope.
Examples of these split profiles are given in Fig.~\ref{fig:splitferm}.  

\begin{figure}
\centering
\includegraphics[width=0.43\textwidth]{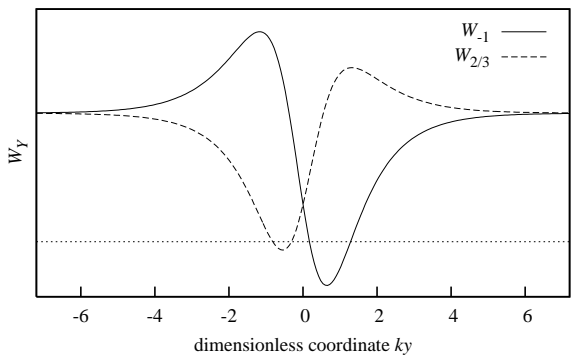}
\caption{
\small
Example potential profiles $W_Y(y)$, Eq.~(\ref{eq:pwcpot}),
which trap the Higgs doublet and coloured scalar.  The straight
horizontal line is $W_Y=0$.  Parameters are chosen such
that the lowest eigenstate of $W_{-1}$ ($W_{2/3}$) has a negative
(positive) eigenvalue. This gives the Higgs doublet a tachyonic
mass on the brane while keeping the coloured scalar heavy.
}
\label{fig:higgspot}
\end{figure}

The Higgs-doublet $\Phi_w$ and coloured-scalar $\Phi_c$ contained in $\Phi$ are similarly
localised~\cite{George} and split by their interaction with the background fields, as
given by the $\lambda_{4-7}$ terms in Eq.~(\ref{eq:Vrest}). Writing
$\Phi_{w,c}(x,y) = p_{w,c}(y) \phi_{w,c}(x)$, where $\phi_{w,c}$
are required to satisfy a massive $3+1$-d Klein-Gordon equation with mass-squared parameters $m^2_{w,c}$,
the profiles $p_{w,c}$ obey the Schr\"{o}dinger-like equation,
\begin{equation}
-\frac{d^2}{dy^2} p_{w,c}(y) + W_Y(y) p_{w,c}(y) = m^2_{w,c} p_{w,c}(y),
\label{eq:pwc}
\end{equation}
with a weak-hypercharge-dependent effective potential,
\begin{equation}
W_Y(y) = \mu_\Phi^2 + \lambda_4 \eta^2 + \lambda_5 \chi_1^2
    + \frac{3 Y^2}{20} \lambda_6 \chi_1^2 + \sqrt{\frac{3}{5}} \frac{Y}{2} \lambda_7 \eta \chi_1.
\label{eq:pwcpot}
\end{equation}
The full spectrum of localised and 
delocalised $\Phi$ modes is obtained by solving these eigenvalue equations, but we are interested
here in only the lowest mass eigenstates.  There is sufficient parameter freedom to
allow $m^2_w < 0$ while $m^2_c >0$, thus setting the stage for an effective Mexican-hat potential for $\phi_w$
and hence electroweak symmetry breakdown inside the wall.  
An example of the effective potentials $W_Y(y)$ are given in Fig.~\ref{fig:higgspot}. (The scalar spectrum
also contains the kink translational zero mode; Ref.~\cite{George} explains how this 
mode can be frozen out.)

We can now see how natural resolutions arise to some of the usual problems with an
SU(5) GUT.  The mass relation $m_e = m_d$ is {\it not} obtained, because the
$3+1$-d Yukawa couplings depend on overlap integrals in the extra dimension, which
will be different because of the different fermion localisation profiles.
The coloured scalar $\phi_c$ induces $p \to \pi^0 e^+$ proton
decay through the Yukawa terms $\overline{u}_R (e_R)^c \phi_c^*$ and $\overline{d}_R (u_R)^c \phi_c$,
but this effect can be suppressed by making the relevant profile overlaps very small.  For example,
splitting $u_R$ and $d_R$ so that that they overlap exponentially little would suffice~\cite{Coulthurst}.

For the one-family standard model, it is obvious that we have enough parameters to fit the quark and lepton
masses.  For the three-family case, it is plausible that sufficient parameter freedom exists, though this
has not been rigorously proven as yet.  It is a complicated problem, because the physical observables
depend on profile functions which depend in complicated ways on the Lagrangian parameters (and corrections
to the classical calculations due to the effect of the non-perturbative bulk will also exist at some level).

Gauge coupling constant evolution cannot be examined until a proper phenomenological parameter fitting is
done, because the higher mass modes both depend on these parameters and affect
the coupling constant evolution.  Since the higher mass modes are split $SU(5)$ multiplets, the running will 
be different from standard $3+1$-d non-supersymmetric $SU(5)$, and successful unification may be possible.
Note that coupling constants run logarithmically, not through a power-law,
in the effective $3+1$-d theory of localised fields.

We now turn on gravity, with Eq.~(\ref{eq:S}) modified to
\begin{equation}
S = \int d^5x \sqrt{G} \left( -2 M^3 R - \Lambda + T - Y_{DW} - Y_5 - V \right),
\end{equation}
where $G$ is the determinant of the metric, $M$ the 5D gravitational mass scale, $R$ the scalar curvature,
and $\Lambda$ the bulk cosmological constant.  The other terms now include minimal coupling to gravity.
We first seek a background $\eta$-$\chi$-metric configuration that will simultaneously localise gauge
bosons and gravitons.  For a significant parameter-space region, the Einstein-Klein-Gordon equations 
admit numerical solutions where $\eta$ is a
kink, $\chi_1$ is an even function that asymptotes to zero at $|y| = \infty$, and the metric
assumes the Minkowski-brane warped form,
\begin{equation}
ds^2_5 = e^{-\rho(y)/6M^3} \eta_{\mu\nu} dx^{\mu} dx^{\nu} - dy^2,
\end{equation}
with $\rho(y) \sim |y|$ asymptotically. The usual Randall-Sundrum fine-tuning condition
involving the bulk cosmological constant must be imposed to ensure a Minkowski brane.
For the special parameter choices,
\begin{eqnarray}
0 & = & 2c - 4l - \tilde{\lambda},\quad a = 0, \nonumber\\
\mu_\chi^2 & = & l v^2 \left ( \frac{6 M^3}{6 M^3 + v^2} \right )
	+ \frac{\tilde{\lambda} v^2}{2} \left ( \frac{9 M^3 + 2 v^2}{6 M^3 + v^2}
	   \right )
\end{eqnarray}
an analytic solution exists; this is useful because it serves as a concrete example:
\begin{eqnarray}
\eta(y) & = & v \tanh(k y), \\
\chi_1(y) & = & v\, {\rm sech}(k y), \\
\rho(y) & = & v^2 \log \left [ \cosh(k y) \right ],
\label{eq:DSRS2}
\end{eqnarray}
where $k^2 = 3 M^3 (c v^2 - \mu_\chi^2 )/(3 M^3 + v^2 )$. As in RS2, the linearised graviton fluctuation equation
has a confined zero-mode that is identified as the usual graviton~\cite{Csaki}.

The $3+1$-d fermion spectrum still contains a localised zero mode for each
species.  However, far from the brane the effective potentials that replace those in Eq.~\ref{eq:fermioneffectivepotential}
are now driven asymptotically to zero by the exponentially decreasing warp factor~\cite{DRT1},
whereas in the gravity-free case they tended to the strictly positive constants $b^2_{nY}(y = \pm\infty)$.  The
gravity-case effective potentials are thus volcano-like and consequently support modes of arbitrarily small energy: 
continua starting at zero mass.  This feature is quite analogous to the well-known graviton-mode situation
in the RS2 model: there is no mass gap, but rather a continuum a modes starting immediately above
the localised massless graviton mode~\cite{RS2}.  
The $\Phi$ spectrum similarly contains a localised standard-model Higgs doublet plus
a continuum starting at zero mass~\cite{DavGeo}.  We need to make sure that the absence of a mass gap
does not spoil the existence of a low-energy effective theory displaying dimensional reduction down to $3+1$-d.

Let us first for simplicity ignore the Dvali-Shifman-like bulk physics, or focus, if you like, on the modes of the gauge-singlet
scalar field $\eta$.  Except for discrete
resonant modes, corresponding to quasi-localised states~\cite{DRT1,DavGeo}, 
the lowest-mass continuum modes are suppressed on the
brane because they have to tunnel through the potential barrier of the volcano-like effective potential.  
As such, their integrated effects at low energies will be dominated
by the zero modes, just as in the well-known graviton case~\cite{RS2}.  Because of this phenomenon,
the localised modes do indeed form a low energy effective $3 + 1$-d
theory.  A detailed discussion of these matters can be found in Ref.~\cite{DavGeo}.  

The analysis of the gauge non-singlet fermion and scalar modes is affected by the DS phenomenon.
Since the continuum modes penetrate into the bulk, they feel the full effects of the confinement-phase physics
we assume holds there.  We therefore expect the low-mass continuum modes to manifest physically as the constituents of massive
``hadrons'' in the bulk.  But the lowest mass hadrons still have to tunnel through the volcano-like potential barriers
to get inside the domain wall, and since the non-perturbative effects switch off near the wall, the situation
analysed in the previous paragraph is regained and with some plausibility the same conclusions follow.

Having described the construction of the model, it is now worth surveying the various scales it contains and how they
should relate to each other.  Of the many scales in the model, four need careful consideration: the ultraviolet cutoff $\Lambda_{\rm UV}$,
the SU(5) breaking scale on the brane $\Lambda_{\rm SU(5)} \sim [\chi_1(y=0)]^{2/3}$, the bulk SU(5) confinement scale $\Lambda_{\rm conf}$
and the DW inverse width $\Lambda_{\rm DW} \equiv k$.  All of these scales must be well above
the electroweak scale.  Within the four, the required hierarchy is
\begin{equation}
\Lambda_{\rm UV} > \Lambda_{\rm SU(5)} > \Lambda_{\rm conf} > \Lambda_{\rm DW}.
\label{eq:scales}
\end{equation}
For obvious reasons, the UV cut-off must be the highest scale in the theory.  The SU(5) breaking scale on the brane must be higher than
the SU(5) bulk confinement scale, because we need to suppress the SU(5) confinement dynamics on the brane.  If the opposite were the case,
then the dynamics of the field $\chi$ would be everywhere dominated by the strong SU(5) interactions and our classical background
scalar field configuration would have no physical relevance.  Finally, the SU(5) bulk glueball radius scale must be smaller than the
width of the DW in order for the Dvali-Shifman effect to work, as discussed in the lattice gauge analysis of Ref.\cite{Laine}.  This
translates into the confinement scale being higher than the inverse wall width.  The UV, DW-width and SU(5) breaking scales are
governed by free parameters, so the required hierarchy amongst those three can always be achieved.  The SU(5) confinement scale
is in principle to be calculated from the UV-cutoff bulk SU(5) gauge theory, and will depend on $\Lambda_{\rm UV}$ and the
dimensionful gauge coupling constant $g$.  If the qualitative behaviour of the pure Yang-Mills theory discussed in Sec.~\ref{sec:DS}
also holds for the complete theory, then we expect there to be a critical coupling $g_c(\Lambda_{\rm UV})$ above which the
theory is confining.  The hypothetical lattice gauge theory calculation would have to allow values of $g > g_c$ to furnish
a $\Lambda_{\rm conf}$ that obeyed Eq.~(\ref{eq:scales}).  This calculation has not been performed.

\section{Conclusion}

In summary, we have proposed a candidate $4+1$-d action for realising a SM-like theory plus gravity dynamically
localised to a domain wall. The dynamical localisation mechanisms for fermions, scalars and gravitons are well understood,
whereas gauge boson localisation is postulated by way of the Dvali-Shifman mechanism.  The DS mechanism is at this stage a
conjecture in the $4+1$-d context because of an incomplete understanding of confinement.  What we have shown is that it is
quite straightforward to construct a DW-localised SM if confinement exists for an SU(5) gauge theory bulk.

The proposed model -- a $4+1$-d SU(5) gauge theory minimally coupled
to gravity -- enjoys some interesting qualitative features.  Notably, the usual
tree-level SU(5) relation $m_d = m_e$ is automatically absent and coloured-Higgs-induced 
proton decay can be suppressed. 

There are a number of open problems, including the following:
\begin{itemize}
\item the veracity of the Dvali-Shifman mechanism in $4+1$-d, as discussed above;
\item to understand the phenomenological implications, including for proton decay, of the gauge bosons that are massive inside the domain wall;
\item to see whether there is enough parameter freedom to fit the three-family standard-model masses and mixing angles while 
obeying experimental bounds on proton decay;
\item to study how the effective $3+1$-d SU(3)$\otimes$SU(2)$\otimes$U(1) gauge coupling constants unify into a $4+1$-d SU(5)
gauge coupling constant;
\item to generate nonzero neutrino masses;
\item to understand the phenomenology of the kink translational zero mode in the gravity case~\cite{ShapZeroMode}.
\end{itemize}

The Dvali-Shifman gauge boson localisation mechanism appears to be a keystone.  If it can work in $4+1$-d, then a whole world of domain-wall
brane model building is opened up, of which the theory presented above is but an example.  If it does not work, then it is not
at all clear that realistic field-theoretic domain-wall brane models exist when the extra dimension is non-compact.
We hope that our efforts lead
to renewed interest in the issue of confinement in higher-dimensional gauge theories.

\acknowledgments{We thank A. Kobakhidze and M. Trodden for useful discussions.  DPG would like to thank
Simon Catterall for discussions about confinement in $4+1$ dimensions and for making available some relevant
lattice gauge theory software.  DPG would also like to thank Martin Schmaltz for a useful discussion
about the scales in the model.  This work was supported
by the Australian Research Council, the Puzey Bequest and the Henry \& Louisa Williams Bequest 
to the University of Melbourne.}


\bibliography{references}

\end{document}